\documentclass[twocolumn, preprintnumbers, superscriptaddress,
  prl]{revtex4-1}
\pdfoutput=1
\usepackage[pdftex]{graphicx}
\usepackage{amssymb,amsmath}
\usepackage[colorlinks,allcolors=blue,linktocpage]{hyperref}

\begin{document}
\preprint{INR-TH-2018-005}
\title{Gravitational Bose-Einstein condensation in the kinetic regime}

\author{D.G. Levkov}\email{levkov@ms2.inr.ac.ru}
\affiliation{Institute for Nuclear Research of the Russian Academy
  of Sciences, Moscow 117312, Russia}
\affiliation{Moscow Institute of Physics and Technology, 
  Dolgoprudny 141700, Russia}
\author{A.G. Panin}
\affiliation{Institute for Nuclear Research of the Russian Academy
  of Sciences, Moscow 117312, Russia}
\affiliation{Moscow Institute of Physics and Technology, 
  Dolgoprudny 141700, Russia}
\author{I.I. Tkachev}
\affiliation{Institute for Nuclear Research of the Russian Academy
  of Sciences, Moscow 117312, Russia}
  \affiliation{Novosibirsk State University, Novosibirsk 630090, Russia}

\begin{abstract}
  We study Bose-Einstein condensation and formation of Bose stars in
  the   virialized dark matter halos/miniclusters by universal
  gravitational interactions. We prove that this phenomenon does occur
  and it is described by kinetic equation. We give expression for the
  condensation time. Our results suggest that Bose stars may form
  kinetically in the mainstream dark matter models such as invisible
  QCD axions and Fuzzy Dark Matter.
\end{abstract}

\maketitle

\paragraph{1. Introduction.}
\label{sec:1.-introduction}
Bose stars are lumps of Bose -Einstein condensate bounded by
self-gravity~\cite{Ruffini:1969qy, Tkachev:1986tr}. They can be made
of condensed dark matter (DM) bosons~--- say, invisible QCD
axions~\cite{Sikivie:2006ni} or Fuzzy DM~\cite{Ringwald:2012hr}. That
is why their physics, phenomenology 
and observational signatures remain in the focus of 
cosmological research for decades~\cite{Schunck:2003kk}, see
recent papers~\cite{Dev:2016hxv, Levkov:2016rkk}. Unfortunately,
formation of Bose stars is still poorly understood and many recent
works have to assume their existence.

In this Letter we study Bose-Einstein condensation in the virialized DM
halos/miniclusters caused by universal gravitational interactions. We work at
large occupation numbers  which is correct if the DM bosons are light. 
Notably, we
consider kinetic regime where the initial coherence length and period
of the DM particles are close to the de Broglie values $(mv)^{-1}$
and $(mv^2)^{-1}$ and much smaller than the halo size $R$ and
condensation time~$\tau_{gr}$,
\begin{equation} 
  \label{eq:tau}
  m v R \gg 1 \;,\qquad\qquad  m v^2 \tau_{gr} \gg 1\;.
\end{equation}
We numerically solve microscopic equations for  the ensemble of
gravitating bosons in this case and find that the Bose stars indeed
form. We derive expression for $\tau_{gr}$ and study kinetics of the
process.

Up to our knowledge, gravitational Bose-Einstein condensation in
kinetic regime has not been observed in simulations before. Old works
considered only contact interactions between the DM
bosons~\cite{Tkachev:1991ka} which are non-universal and suppressed by
quartic constants ${\lambda   \sim 10^{-50}}$~\cite{Klaer:2017ond} and
$10^{-100}$~\cite{Arvanitaki:2009fg} in models of QCD axions   and
string axions/Fuzzy DM. Our results show that in these cases
gravitational condensation is {\it faster}: although the  Newton's
constant $Gm^2$ is tiny, its effect is enhanced by collective
interaction of large fluctuations in the boson gas at large distances,
cf.~\cite{Sikivie:2009qn}.

On the other hand, all previous numerical studies of Bose star formation 
considered coherent initial configurations of the bosonic field~---
a Gaussian wavepacket~\cite{Seidel:1993zk} or the Bose stars 
themself~\cite{Schive:2014hza, Schwabe:2016rze}. A spectacular
simulation of structure formation by wavelike/Fuzzy
DM~\cite{Schive:2014dra, Schive:2014hza}
started from (almost) homogeneous Bose-Einstein condensate.
In all these cases the Bose stars form almost
immediately~\cite{Seidel:1993zk, Schive:2014hza} from the
lowest-energy part of the initial condensate.

\begin{figure}[b!]
   \unitlength=1mm
    \includegraphics[width=\columnwidth]{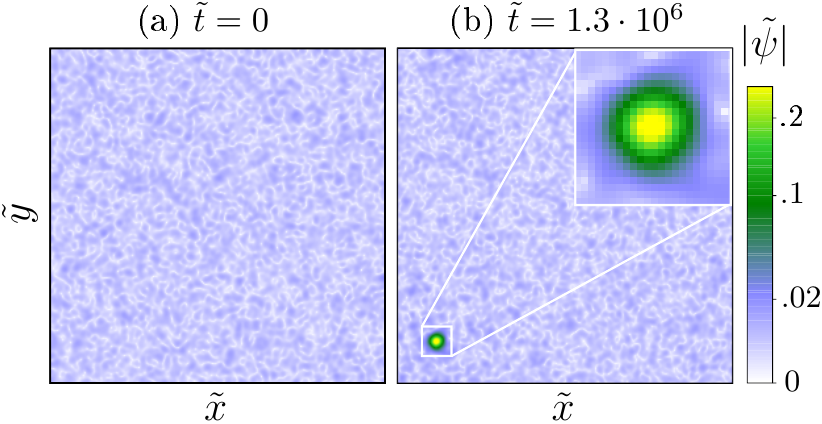}\\ [1ex]
    \hspace{-1.5mm}
    \begin{picture}(86.4,30)
      \put(1.5,0){\includegraphics[width=3.4cm]{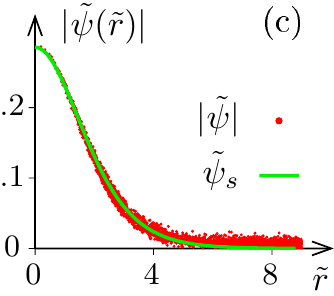}\hspace{1mm}
        \includegraphics[width=4.6cm]{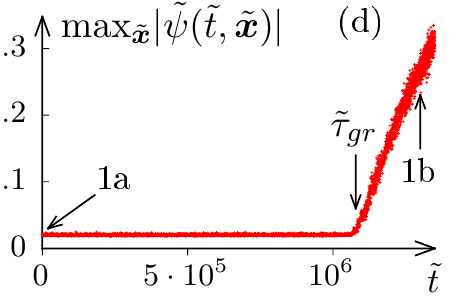}}
    \end{picture}
    \begin{picture}(86.4,35)
      \put(0,0){\includegraphics[width=8.15cm]{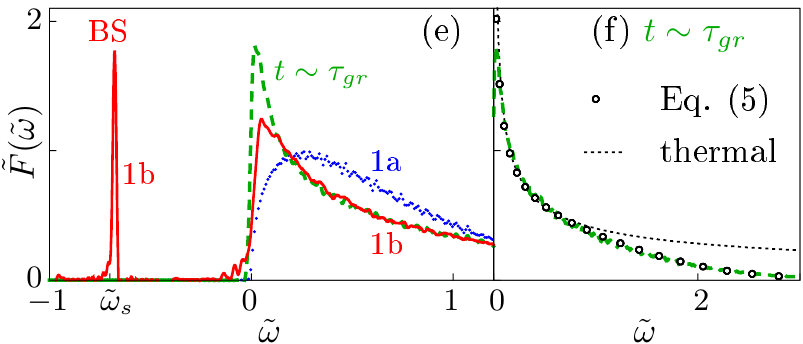}}
    \end{picture}
    \caption{Formation of Bose star from random  field with 
      initial distribution $|\tilde{\psi}_{\tilde{\boldsymbol{p}}}|^2
      \propto \mathrm{e}^{-\tilde{\boldsymbol{p}}^2}$ and total mass
      ${\tilde{N}=50}$  in the box ${0  \leq \tilde{x},\, \tilde{y},\, \tilde{z} < 125}$. 
      These values correspond to the center of  the axion minicluster
      with $M_c\sim 10^{-13} M_\odot$ and $\Phi \sim 2.7$ in Sec. 8.
    (a),~(b)~Sections
    $\tilde{z}=\mbox{const}$ of the solution $|\tilde{\psi}(\tilde{t},\,
    \tilde{\boldsymbol{x}})|$ at (a) $\tilde{t}=0$ and (b)~$\tilde{t}
    > \tilde{\tau}_{gr}  \approx 1.08 \cdot 10^6$. (c)~Radial profile
    $|\tilde{\psi}(\tilde{r})|$ of the object in
    Fig.~\ref{fig:sections1}b (points) compared to the Bose star
    $\tilde{\psi}_s(\tilde{r})$ with $\tilde{\omega}_s\approx -0.7$
    (line). (d)~Maximum  of
    $|\tilde{\psi}(\tilde{\boldsymbol{x}})|$ over the box as a function  
    of  time. (e) Spectra (\ref{eq:1}) at times of
    Figs.~\ref{fig:sections1}a,~b and at the eve  of Bose star
    nucleation, $\tilde{t} = {1.05\cdot 10^6}\sim \tilde{\tau}_{gr}$. 
    (f)~The spectrum at $t\sim \tau_{gr}$ (dashed line)
    versus  the solution of Eq.~(\ref{eq:2})
    (circles) and thermal law $\tilde{F} \propto
    \tilde{\omega}^{-1/2}$~(dots). \label{fig:sections1}} 
\end{figure}
We consider entirely different situation~(\ref{eq:tau}) when the DM
bosons are virialized in the initial state. The closest study to
  ours was performed in~\cite{Khlebnikov:1999pt}, but kinetic regime
  was not achived there due to computational limitations.  Note that
  we do not consider scenario of~\cite{Sikivie:2009qn}, where axions
  form cosmological condensate at the radiation-dominated stage,
  cf.~\cite{Guth:2014hsa}.  Indeed, at realignment the momenta of
  such axions are comparable to the Hubble scale, and
  Eq.~(\ref{eq:tau}) is violated.

\paragraph{2. The birth of the Bose star.} 
\label{sec:2.-bose-star}
Consider $N$ nonrelativistic
gravitationally interacting bosons in  the 
periodic box of size $L$. At large occupation numbers this system is
described by a random classical field $\psi(t,\,
\boldsymbol{x})$~\cite{Tkachev:1991ka} evolving in its own
gravitational potential~$U(t,\, \boldsymbol{x})$, 
\begin{align}
  \label{eq:GP2}
  &  i \partial_t \psi = -\Delta \psi/2m  + mU \psi \;, \\
  \notag
  & \Delta U = 4\pi G\, m (|\psi|^2-n) \;,
\end{align}
where the mean particle density ${n \equiv N/L^3}$  is subtracted in the
second line for consistency~\cite{Schive:2014dra}. Notably,
Eqs.~(\ref{eq:GP2}) simplify in dimensionless variables: 
substitutions $\boldsymbol{x} =   \tilde{\boldsymbol{x}}/ mv_0$, $t =
\tilde{t}/mv_0^2$, $U = v_0^2\tilde{U}$ and ${\psi =
  v_0^2\tilde{\psi}\sqrt{m/G}}$ exclude parameters $m$ and $G$ 
from the equations and reference velocity $v_0$~--- from the initial
conditions. The rescaled particle number is $\tilde{N} \equiv
\int d^3 \tilde{\boldsymbol{x}} \, |\tilde{\psi}|^2= Gm^2 N/v_0$. 

We fix initial conditions in the momentum space.  A
  representative class of them describes Gaussian-distributed bosons, 
${|\tilde{\psi}_{\tilde{\boldsymbol{p}}}|^2   = 
  8\pi^{3/2}\tilde{N} \, 
  \mathrm{e}^{-\boldsymbol{\tilde{p}}^2}}$,
with  random  phases
$\mbox{arg}\,\tilde{\psi}_{\tilde{\boldsymbol{p}}}$, where $\tilde{\boldsymbol{p}} \equiv
\boldsymbol{p}/mv_0$. Fourier-transforming  
$\tilde{\psi}_{\tilde{\boldsymbol{p}}}$, we obtain an isotropic and
homogeneous initial configuration $\tilde{\psi}(0,\,
\tilde{\boldsymbol{x}})$ with  minimal coherence length in
Fig.~\ref{fig:sections1}a. Then 
we numerically evolve Eqs.~(\ref{eq:GP2}) using an  exceptionally
stable $3D$ algorithm,  see Supplemental material~S2 and 
movie~\cite{movie_box}. Apart from the erratic  motion of $\psi$-peaks
and  deeps, nothing happens for a long time $t<\tau_{gr}$, where
$\tilde{\tau}_{gr}  \sim 10^6$ for the solution in 
Fig.~\ref{fig:sections1}. Then a  coherent, compact and
spherically-symmetric object appears at $t>\tau_{gr}$, see
Fig.~\ref{fig:sections1}b. With time the object grows in mass and
moves in a Brownian way due to interaction with the fluctuating
environment.

To explain what happens, we recall that any interaction between the
bosons should lead to thermal equilibrium, and
in the case of  large occupation numbers~---
to formation of a Bose-Einstein condensate. Gravitational
interaction is not an exception,   as pioneering
works~\cite{Khlebnikov:1999pt, Sikivie:2009qn} argued.
But then the condensate cannot appear in a homogeneous
state~\cite{Guth:2014hsa}. Rather, it should  fragment due to Jeans
instability into a set of isolated Bose stars,
cf.~\cite{Tkachev:1991ka}, which is therefore  the actual end-state
of the condensation process. 

The field profiles of the isolated Bose stars are found by solving
Eqs.~(\ref{eq:GP2}) with the  stationary spherical Ansatz ${\psi
  = \psi_s(r) \,  \mathrm{e}^{-i\omega_s t}}$, $U = U_s(r)$ at each $\omega_s<0$,
see e.g.~\cite{Chavanis:2011zi}. We computed them using a separate
code. Results coincide with the profiles of condensed
objects on the lattice (see Fig.~\ref{fig:sections1}c) thus proving that
we indeed observe nucleation of Bose stars. We performed 
simulations for a large set of parameters, for $\delta$- and
$\theta$-like initial distributions, $|\psi_{\boldsymbol{p}}|^2 \propto
\delta(|\boldsymbol{p}|-mv_0)$ and  $\theta(mv_0 -
|\boldsymbol{p}|)$, in addition to the Gaussian. Every time we
observed formation of 
a Bose star with correct profile  $\psi_s(r)$, $U_s(r)$ correct
  mass $M_s \propto \psi_s^{1/2}(0)$ and correct size proportional
  to $M_s^{-1}$, see~\cite{Chavanis:2011zi}.  Note that the Bose stars
nucleate wide and 
rarefied, then shrink and become dense as they accumulate bosons.  
Unlike in other studies, no ``seed'' Bose condensate was present in
our simulations at $\tau< \tau_{gr}$, otherwise it would grow above the
background in a short time, see Fig.~\ref{fig:sections1}d.  

\paragraph{3. The spectrum.}
To look deeper into the initial, seemingly featureless stage of gas
evolution, we compute distribution $F(t,\,\omega) = dN/d\omega$ of
bosons over energies $\omega$. This quantity equals to Fourier image
of the correlator
\begin{equation}
  \label{eq:1}
  F = \int \frac{dt_1}{2\pi} \, d^3\boldsymbol{x} \, \psi^*(t,
  \boldsymbol{x}) \psi(t+t_1,\boldsymbol{x})\,  \mathrm{e}^{i\omega 
  t_1 - t_1^2/\tau_1^2 }
\end{equation}
in kinetic regime $(m v_0^2)^{-1}   \ll \tau_1 \ll
\tau_{gr}$,  see Supplementary material S1
  and~\cite{Zakharov-Lvov}. In dimensionless calculations we use 
${\tilde{F} = mv_0^2 F/N}$ normalized to unity: $\int  \tilde{F}\,
d\tilde{\omega}  =  1$, where $\tilde{\omega} = \omega / 
mv_0^2$. 

Figure \ref{fig:sections1}e shows that the spectrum (\ref{eq:1})
completely changes during evolution at ${t<\tau_{gr}}$. It starts from
a  wide bell ${\tilde{F}\propto   \tilde{\omega}^{1/2} \,
  \mathrm{e}^{-2\tilde{\omega}}}$ corresponding to Gaussian
distribution in momenta in Fig.~\ref{fig:sections1}a. As the time goes
on, $F$ develops a peak at low $\omega$ and becomes close
to  thermal at intermediate energies, $F \propto \omega^{-1/2}$, 
see the graph at $t\sim  \tau_{gr}$ in Figs.~\ref{fig:sections1}e and
\ref{fig:sections1}f.  At high $\omega$ it still falls off
  exponentially, as high-frequency modes thermalize slowly~\cite{Zakharov}.
Once the Bose star nucleates, a small $\delta$-peak appears in the
distribution.  With time this $\delta$-peak grows in height and
  shifts to the left, see the spectrum 1b in
  Fig.~\ref{fig:sections1}e.  It   explicitly shows condensed
particles of  the same energy $\omega_s<0$ inside the 
  growing Bose star. 

Below we use the $\delta$-peak at $\omega<0$ as an indicator of Bose star
nucleation: we define $\tau_{gr}$ as the moment when the peak is twice
higher than the fluctuations in $F(t,\, \omega)$.  

\paragraph{4. Condensation time.}
\label{sec:4.-cond-time}
In kinetic regime evolution of $F(t,\, \omega)$ is described by
kinetic equation~--- this fact can be proven by
solving Eqs.~(\ref{eq:GP2}) perturbatively and using
approximations~(\ref{eq:tau}),  see Supplemental material S1 and
cf.~\cite{Zakharov-Lvov}. One therefore expects that the time 
of Bose star formation $\tau_{gr}$ is proportional with some
coefficient $b$ to the kinetic relaxation time: ${\tau_{gr} =
  4b\sqrt{2}/ (\sigma_{gr} v n\, f})$, where  $\sigma_{gr} \approx 8\pi (mG)^2
\Lambda/ v^4$  is the transport Rutherford cross section of
gravitational scattering, $\Lambda = \log(mvL)$ is the Coulomb
logarithm, and $f = 6\pi^2 n/(mv)^3 \gg 1$ is the phase-space density
that appears due to Bose stimulation~\cite{Tkachev:1991ka}. The
coefficient $b=O(1)$  accounts for the details of the process. It is
expected to depend weakly on the initial distribution.

Taking all factors together, we obtain expression 
\begin{equation}
\label{eq:tgra3}
\tau_{gr} = \frac{b\sqrt{2}}{12\pi^3}\,  \frac{m  v^6}{G^2 n^2
  \Lambda}\;, \qquad b\sim 1\;,
\end{equation}
that apart from the Coulomb logarithm  $\Lambda$ involves only local parameters
i.e.\ the boson number density $n$ and characteristic velocity~$v$. So, up to weak
logarithmic dependence on the size $L$ formation of  the Bose star can be 
regarded as a local process, with periodic box representing a central
part of some DM halo. We will confirm this intuition below.  

\begin{figure}
  \unitlength=1mm
   \begin{picture}(86.4,35)
     \put(0.2,0){\includegraphics[width=83.5mm]{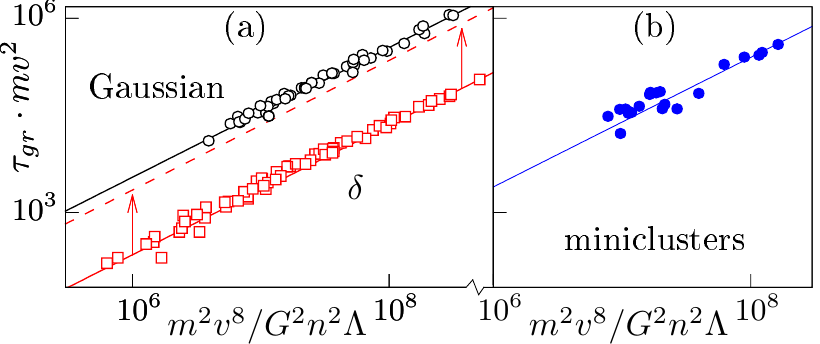}}
   \end{picture}\\[2ex]
   \includegraphics[width=\columnwidth]{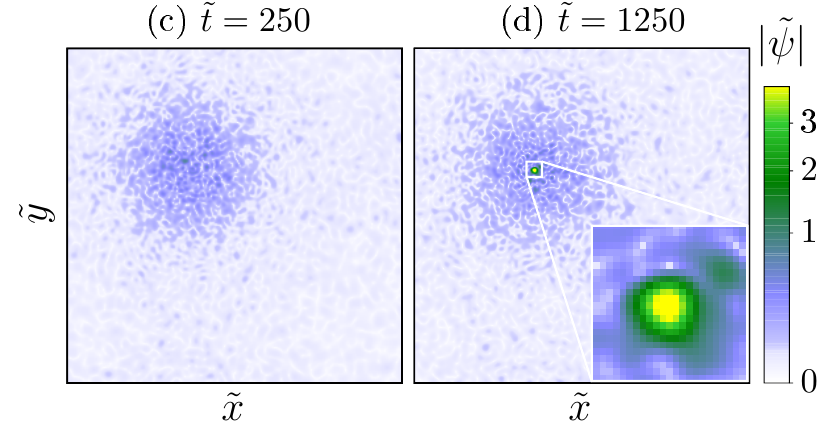}
   \caption{(a) Time to Bose star formation in the cases
     of Gaussian (\begin{minipage}{1.5mm} 
       \protect\includegraphics{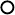}\end{minipage}) and
     $\delta$-peaked (\begin{minipage}{1.5mm}
       \protect\includegraphics{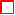}\end{minipage}) initial
     distributions. The 
     $\delta$-graphs are shifted downwards ($\tau_{gr} \to
     \tau_{gr}/10$) for visualization purposes. Lines depict fits by
     Eq.~(\ref{eq:tgra3}). (b)~The same for isolated 
     miniclusters. (c),~(d)~Slices $\tilde{z} = \mbox{const}$ of the 
     solution $|\tilde{\psi}(\tilde{t},\, \tilde{\boldsymbol{x}})|$
     describing formation of a Bose star   
     in the center of a minicluster; $\tilde{N}=290$,
    $\tilde{L}\approx 63$.\label{fig:tgra}}
\end{figure}

We performed simulations of the gas with Gaussian initial distribution
at different $\tilde{L}$ and $\tilde{n}$. Our
results for $\tau_{gr}$ (circles in Fig.~\ref{fig:tgra}a) cover two
orders of magnitude, but they are nevertheless well
fitted by Eq.~(\ref{eq:tgra3}) with $v=v_0$ and $b\approx 0.9$ (upper
line in Fig.~\ref{fig:tgra}a).  To confirm that  Eq.~(\ref{eq:tgra3})
is universal, we repeated the calculations for the 
initial $\delta$-distribution, $|\psi_{\boldsymbol{p}}|^2 \propto 
\delta(|\boldsymbol{p}| - mv_0)$ (squares in
Fig.~\ref{fig:tgra}a). The new vales of $\tau_{gr}$ are still
described by Eq.~(\ref{eq:tgra3}), albeit with slightly different  
coefficient $b\approx 0.6$.
We conclude that Eq.~(\ref{eq:tgra3}) is a practical  and
justified expression for the time of Bose star formation.

\paragraph{5. Kinetics.} Let us show that evolution of $F(t,\,
\omega)$ in Fig.~\ref{fig:sections1}e is indeed governed by the Landau
kinetic equation~\cite{Zakharov} for  the homogeneous isotropic
  Coulomb ensemble,
\begin{equation}
  \label{eq:2}
  \partial_t \tilde{F} = \tau_0^{-1} \partial_{\tilde{\omega}} \left[
    \tilde{A}  \partial_{\tilde{\omega}} \tilde{F}  +
    (\tilde{B} \tilde{F} - \tilde{A}) 
    \tilde{F}/2\tilde{\omega}\right]\;,
\end{equation}
see Supplementary material S1.4 for derivation. 
Here the scattering integral in the right-hand side involves
$\tilde{A}(\tilde{\omega}) = \int_0^{\infty} d\tilde{\omega}_1   
\,\mathrm{min}^{3/2}(\tilde{\omega},\, \tilde{\omega}_1)
\tilde{F}^2(\tilde{\omega}_1)/  (3 \tilde{\omega}_1
\tilde{\omega}^{1/2})$,
$\tilde{B}(\tilde{\omega}) = \int_0^{\tilde{\omega}} d\tilde{\omega}_1
\tilde{F}(\tilde{\omega}_1)$, it is explicitly proportional to the
inverse relaxation time  $\tau_0^{-1} =  8\pi^3 n^2 G^2 (\Lambda + a)/mv_0^6
\sim \tau_{gr}^{-1}$. Notably,  Eq.~(\ref{eq:2}) is valid in the
leading logarithmic approximation ${\Lambda \gg 1}$ which is too rough
for our numerical solutions with $\Lambda \sim 5$. To get a
quantitative comparison, we added an unknown correction $a = O(1)$
to~$\Lambda$.   

We numerically evolve Eq.~(\ref{eq:2}) starting from the same initial
distribution as in Fig.~\ref{fig:sections1}. In
Fig.~\ref{fig:sections1}f  the solution $F(t,\,\omega)$  (circles)
is compared to the microscopic distribution
(\ref{eq:1}) (dashed line) at $t \approx \tau_{gr}$, where $a \approx
5$ is obtained from the fit. We observe agreement in the kinetic
region $\tilde{\omega} \gg 2\pi^2/\tilde{L}^2$ which  confirms 
Eq.~(\ref{eq:2}) at $t< \tau_{gr}$.

Note that unlike in the case of short-range
interactions~\cite{Semikoz:1994zp} thermalization in Landau equation
does not proceed via  power-law turbulent 
cascades~\cite{Zakharov}, and we do not observe them   
in Figs.~\ref{fig:sections1}e,f. Nevertheless, we think that 
Eq.~(\ref{eq:2}) provides the basis for analytic description  
of gravitational Bose-Einstein condensation.

\paragraph{6. Miniclusters.}
\label{sec:6.-miniclusters}
So far we assumed that homogeneous  ensemble in the box correctly
describes central parts of DM halos. Now, we study the isolated 
halos/miniclusters themselves and verify this assumption. Recall that
in large  volume nonrelativistic gas  forms clumps at scales $R \gtrsim
2\pi / k_J$ 
due to Jeans instability, where ${k_J^2 = 2\pi G n m^2\langle
  \omega^{-1} \rangle}$ and the average is computed with
$F(\omega)$.  Starting numerical evolution from the homogeneous  ensemble
with $\delta$-distributed momenta at ${L > 2\pi / k_J}$, we indeed 
observe formation of a virialized minicluster in
Fig.~\ref{fig:tgra}c. With time it remains stationary until  the Bose
star appears in its center, see Fig.~\ref{fig:tgra}d and
movie~\cite{movie_box}. Thus,  formation of Bose stars is not specific 
to finite boxes.  

We checked that our kinetic expression for $\tau_{gr}$ works for the
virialized miniclusters. To this end we generated many different  
miniclusters, computed their central densities $n$ and virial velocities
$\langle v^2 \rangle = -2 \langle \omega \rangle /m$ using the
${\omega < 0}$ part of the distribution $F(\omega)$,  estimated
their radii $R$. In Fig.~\ref{fig:tgra}b we plot the times of Bose star
formation in the miniclusters versus these parameters and 
$\Lambda = \log(mvR)$ (points). The numerical data are well
approximated by Eq.~(\ref{eq:tgra3}) with $b\approx 0.7$ (line)
although the statistical fluctuations are now larger due to limited
control over momentum distribution inside the miniclusters.

Estimating the virial velocity $v^2 \sim 4\pi GmnR^2/3$ in the halo
of radius $R$, one recasts Eq.~(\ref{eq:tgra3}) in the intuitively
simple form ${\tau_{gr} \sim 0.047\, (R/v) \, (Rmv)^3/\Lambda}$,
where the numerical factor is computed. Remarkably, $\tau_{gr}$
equals to the free-fall time $R/v$ multiplied by the cube of kinetic constant
$Rmv \gg 1$ in Eq.~(\ref{eq:tau}). In non-kinetic case $Rmv\sim 1$ the 
Bose stars form immediately~\cite{Seidel:1993zk, Schive:2014hza,
  Schive:2014dra}.

\paragraph{7. Bose star growth.}
\label{sec:7.-bose-star}
After nucleation the Bose stars start to acquire particles from the
 ensemble. Due to computational limitations we are able 
to observe only the first decade of their mass increase that proceeds
according to the heuristic law ${M_s(t) \simeq c v_0
  (t/\tau_{gr}-1)^{1/2}/Gm}$ with  $c = 3 \pm 0.7$. The ratio
$t/\tau_{gr}$ in this expression suggests that growth of the Bose
stars is a kinetic process deserving a separate study.   

\paragraph{8. Discussion.}
\label{sec:7.-discussion}
Let us argue that the Bose stars  appear in the popular cosmological
models  even if we conservatively assume that halos/miniclusters in
  these models are initially virialized. If the DM is made of
invisible QCD  axions~\cite{Sikivie:2006ni}, the  smallest
  substructures are  axion miniclusters~\cite{Hogan:1988mp,
    Kolb:1995bu} of typical mass $M_c \sim 10^{-13}
  M_\odot$~\cite{Enander:2017ogx}. These  miniclusters can be
characterized~\cite{Kolb:1995bu} by the ratio ${\Phi+1 \equiv 
  n/\bar{n}|_{RD}}$  of their central density $n$ to the cosmological 
axion density $\bar{n}$ at the radiation-dominated stage when they are
still in the linear regime. Substituting their typical
  parameters~\cite{Kolb:1995bu} into Eq.~(\ref{eq:tgra3}) and
expressing the result in terms of $\Phi$ and $M_c$, we find,
\begin{equation}
  \notag
        {\tau}_{gr} \sim \frac{10^{9}\, \mbox{yr}}{\Phi^3 (1+\Phi)}
\left(\frac{M_c}{10^{-13}\, M_\odot}\right)^2 \left(\frac{m}{26\, \mu
  \rm eV} \right)^3\;,
\end{equation}
where the reference axion mass is taken
  from~\cite{Klaer:2017ond}. Thus,  typical miniclusters with $\Phi
\sim 1$ condense during the lifetime 
of the Universe, the densest ones with ${\Phi \sim
10^3}$~\cite{Kolb:1995bu}~--- in several hours. The Bose stars are 
important~\cite{Dev:2016hxv} as they hide a part of DM from
observations. After becoming large they  may explode into relativistic
axions~\cite{Levkov:2016rkk} or emit
radiophotons via parametric
resonance~\cite{Tkachev:1986tr} which at different
redshifts may explain FRB~\cite{Tkachev:2014dpa} and anomalies of
ARCADE~2 and EDGES~\cite{Fraser:2018acy}.

Note that gravitational relaxation of  virialized QCD axions is
significantly faster than relaxation due to self-coupling $\lambda \equiv
m^2/f_a^2$, where $f_a \sim 10^{11}\, \mbox{GeV}$ is the Peccei-Quinn
scale. Indeed,  in kinetic regime the relaxation rates are 
proportional to the cross sections, and $\tau_{self}/\tau_{gr}
\sim \sigma_{gr}/\sigma_{self} \sim (10f_aG^{1/2}/v)^4$. In typical
miniclusters $v \sim 10^{-10} \ll 10 f_a G^{1/2}$, and
gravitational interactions win by $\tau_{self} / \tau_{gr} \sim
10^{12}$.

Another popular class of DM models is based on string axions / Fuzzy
DM~\cite{Ringwald:2012hr}. An interesting though recently
constrained~\cite{Irsic:2017yje} scenario considers the smallest mass  
${m\sim 10^{-22} \, \mbox{eV}}$ of these
particles~\cite{Schive:2014dra} when their de Broglie wavelength
inside the dwarf galaxies is comparable to the size of the galaxy
cores, $mv R\sim 1$. As we argued, the Bose stars should appear  in
these cores in free-fall time. This explains their fast
formation~\cite{Schive:2014dra} in numerical  simulations. At larger
masses one substitutes typical  parameters of dwarf satellites into
Eq.~(\ref{eq:tgra3}),
$$
    {\tau}_{gr} \sim 10^{6}\, \mbox{yr}
    \left(\frac{m}{10^{-22}\, \rm eV} \right)^3 
    \left(\frac{v}{30\, \rm km/s} \right)^6
    \left(\frac{0.1 \, M_\odot/ \rm pc^{3}}{\rho}\right)^2.
$$
The Bose stars nucleate there if ${m \lesssim 2 \cdot 10^{-21} \,
  \mbox{eV}}$, at the boundary of experimentally allowed mass
window~\cite{Irsic:2017yje}. Then the missing satellites may hide
as Bose stars. At even larger $m$ the Bose stars may form in miniclusters
and in cores of large galaxies, they may grow overcritical and 
explode~\cite{Levkov:2016rkk}. Note that self-interaction of typical
string axions~\cite{Arvanitaki:2009fg} with  $f_a  \sim 10^{-2}G^{-1/2}$
is less effective than gravity because $v \ll 10f_aG^{1/2}$ in all
structures. 

We thank A.~Pustynnikov, D.~Gorbunov, M.~Ivanov, E.~Nugaev and V. Rubakov 
for discussions. This work was supported by the grant RSF
16-12-10494. Numerical calculations  were performed on the
Computational cluster of TD~INR~RAS.


\onecolumngrid
\clearpage  
\begin{center}
\textbf{\large {\it Supplemental material on the article:}\\
  Gravitational Bose-Einstein condensation in the kinetic regime}
\end{center}

\vspace{5mm}

\twocolumngrid

\setcounter{equation}{0}
\setcounter{figure}{0}
\setcounter{table}{0}
\setcounter{page}{1}
\makeatletter
\renewcommand{\theequation}{S\arabic{equation}}
\renewcommand{\thefigure}{S\arabic{figure}}
\renewcommand{\bibnumfmt}[1]{[S#1]}
\renewcommand{\citenumfont}[1]{S#1}

\section{S1. Landau equation}
\label{sec:s1.-landau-equation}

Evolution of a random Bose field at large occupation numbers can be
well approximated by the classical field theory. Such situations
appear, in particular, in systems far from thermal equilibrium when
particles have energies much lower than the temperature in the
eventual thermal state, $\boldsymbol{p}^2 \ll 2m T$, see
e.g. \cite{Khlebnikov:1996mc}. For the period of evolution with this
property the seminal ``ultraviolet catastrophe'' is not threatening  
the theory and quantum statistics is not needed. This allows us to
describe the initial stages of thermalization by solving the classical
Schr\"odinger-Poisson problem~\eqref{eq:GP2}. In our simulations the
energies of all particles at the epoch of Bose star formation  are
well below the lattice UV cutoff, not to say about the region of
parametrically  high momenta corresponding to  order-one occupation
numbers in the thermal ensemble, see Fig.~\ref{fig:sections1}f. 

Let derive Landau kinetic equation for the
gravitating ensemble of random classical waves \eqref{eq:GP2} inside
the structure of size $R$~--- a halo/minicluster or a periodic box. The
main steps of our derivation are standard~\renewcommand{\citenumfont}[1]{#1}\cite{Zakharov-Lvov}\renewcommand{\citenumfont}[1]{S#1},
although specific attention should be paid to kinetic corrections
suppressed by powers of $(mvR)^{-2}$ and to infrared divergence of
the scattering integral caused by long-range nature of 
gravitational interactions. Treating these subtleties, we justify Landau
approximation for the integral thus supporting the  relaxation time
estimates from the main text.

\subsection{S1.1. Vlasov equation}
We introduce real  Wigner distribution for the ensemble of random
classical fields,
\begin{equation}  
  \label{eq:S2}
  f_{\boldsymbol{p}}(\boldsymbol{x}) = \int d^3 \boldsymbol{\xi} \;
  \mathrm{e}^{-i \boldsymbol{p} \boldsymbol{\xi}}\,  \langle
  \psi_{\boldsymbol{x} + \boldsymbol{\xi}/2} \; \psi^*_{\boldsymbol{x} -
  \boldsymbol{\xi}/2} \rangle\;,
\end{equation}
where $\psi_{\boldsymbol{x}} \equiv \psi(\boldsymbol{x})$ and $\langle
\cdot\rangle$ is the average with respect to random initial phases
of the fields.  In the kinetic regime~\eqref{eq:tau} the value of 
$f_{\boldsymbol{p}}(\boldsymbol{x})$ can be 
interpreted~\renewcommand{\citenumfont}[1]{#1}\cite{Zakharov-Lvov}\renewcommand{\citenumfont}[1]{S#1} as an occupation number of bosons in
the elementary cell $d^3 \boldsymbol{x}d^3\boldsymbol{p} = 
(2\pi)^3$.

Taking the time derivative of Eq.~(\ref{eq:S2}) and using
Eqs.~\eqref{eq:GP2},  one obtains dynamical equation for
$f_{\boldsymbol{p}}(\boldsymbol{x})$,
\begin{multline}
  \label{eq:S1}
  \partial_t f + \boldsymbol{p}\boldsymbol{\nabla}_x f/m \\=  2 m\,
  \mathrm{Im}\int d^3 \boldsymbol{\xi}\;    \mathrm{e}^{-i
    \boldsymbol{p}  \boldsymbol{\xi}}\,\langle
  \psi_{\boldsymbol{x}+\boldsymbol{\xi}/2} \psi^*_{\boldsymbol{x} -
    \boldsymbol{\xi}/2} U_{\boldsymbol{x} +
    \boldsymbol{\xi}/2}\rangle\;,
\end{multline}
where $\mathrm{Im}$ extracts $\boldsymbol{\xi}$-odd part of
$U_{\boldsymbol{x}+\boldsymbol{\xi}/2}$. Note that the gravitational 
potential equals
\begin{equation}
  \label{eq:S4}
  U_{\boldsymbol{x}} = 4\pi G m \int d^3\boldsymbol{x}'\,
  \Delta^{-1}_{\boldsymbol{x}-\boldsymbol{x}'} (|\psi_{\boldsymbol{x}'}|^2-n)\;.
\end{equation}
Here $\Delta^{-1}$ is the Green's function of the Poisson
equation~\eqref{eq:GP2}: $\Delta \Delta^{-1}_{\boldsymbol{x}} \equiv
\delta^{3}(\boldsymbol{x})$ and $\Delta^{-1}_{-\boldsymbol{x}} =
\Delta^{-1}_{\boldsymbol{x}}$. Thus, Eq.~(\ref{eq:S1}) contains a
correlator of four $\psi$-fields. We should express it via  
$f_{\boldsymbol{p}}(\boldsymbol{x})$ in order to obtain kinetic
equation in the closed form.

To this end we exploit weak coupling expansion. Recall that in the
initial state $\psi$ is a Gaussian random field,  all
non-Gaussianities appear later due to small gravitational
interactions. This means that the Wick theorem
remains approximately valid,
\begin{multline}
  \label{eq:Fpoint}
  \langle \psi_{\boldsymbol{x}_1} \psi^*_{\boldsymbol{x}_2}
  \psi_{\boldsymbol{x}_3}
  \psi^*_{\boldsymbol{x}_4}\rangle = \langle
  \psi_{\boldsymbol{x}_1} \psi^*_{\boldsymbol{x}_2} \rangle \langle
  \psi_{\boldsymbol{x}_3} \psi^*_{\boldsymbol{x}_4}\rangle \\+
  \langle
  \psi_{\boldsymbol{x}_1} \psi^*_{\boldsymbol{x}_4} \rangle \langle
  \psi_{\boldsymbol{x}_3} \psi^*_{\boldsymbol{x}_2}\rangle +
    \langle \psi_{\boldsymbol{x}_1} \psi^*_{\boldsymbol{x}_2} 
  \psi_{\boldsymbol{x}_3} \psi^*_{\boldsymbol{x}_4}\rangle_{conn}\;,
\end{multline}
where the connected part $\langle \cdot \rangle_{conn} \sim
O(G)$ is small, it gives $O(G^2)$ contribution into Eq.~(\ref{eq:S1}).   

We substitute Eqs.~\eqref{eq:S4} and~\eqref{eq:Fpoint} into
Eq.~(\ref{eq:S1}) and express all two-point functions via
$f_{\boldsymbol{p}}(\boldsymbol{x})$, Eq.~(\ref{eq:S2}). This gives 
non-local equation
  \begin{multline}
    \label{largeD}
    \partial_t f + \boldsymbol{p} \boldsymbol{\nabla}_xf /m =
            2m \, \mathrm{Im} \,
    \bar{U}(\boldsymbol{x}
    +\mbox{$\frac{i}{2} \boldsymbol{\nabla}_p$}) \,
    f_{\boldsymbol{p}}(\boldsymbol{x}) \\
    + \frac{8 \pi G m^2}{(2\pi)^6}\, \mathrm{Im} \int d^3 \{\boldsymbol{\xi}
       \boldsymbol{\xi}' \boldsymbol{q}  \boldsymbol{q}'\}\;
       \Delta^{-1}_{\boldsymbol{\xi}} \, 
     \mathrm{e}^{i\boldsymbol{\xi}' \boldsymbol{q} - i
       \boldsymbol{\xi} \boldsymbol{q}'} \\ \times
     f_{\boldsymbol{p} + \boldsymbol{q}} (\boldsymbol{x} +
     \mbox{$\frac{\boldsymbol{\xi}}{2}$}) 
     \, f_{\boldsymbol{p}+\boldsymbol{q}'}(\boldsymbol{x}
     +\mbox{$\frac{\boldsymbol{\xi}'}{2}$}) + \mbox{St}\, f\;,
  \end{multline}
where $\bar{U}(\boldsymbol{x}) \equiv \langle
U(\boldsymbol{x})\rangle$ is the mean potential satisfying
\begin{equation}
  \label{eq:Uav}
  \Delta \bar{U} = 4\pi G m \left( \int \frac{d^3 \boldsymbol{p}}{(2\pi)^3} \,
  f_{\boldsymbol{p}} (\boldsymbol{x}) - n\right)\;,
\end{equation}
and scattering integral $\mbox{St}\, f \sim O(G^2)$ represents
contribution of the connected correlator in
Eq.~(\ref{eq:Fpoint}).  

Ignoring $\mbox{St}\, f$ in Eq.~(\ref{largeD}), one can get an
equation for $f_{\boldsymbol{p}}(\boldsymbol{x})$ valid to the first
order in $G$ and to all orders in the kinetic parameter~$(mvR)^{-1}$. This
simplification is unacceptable for us,  however. First,  all terms in Eq.~(\ref{largeD}) 
except for $\mathrm{St}\, f$  are $T$-odd: they
change sign under the time reflection
$f_{\boldsymbol{p}}(\boldsymbol{x}) \to f_{-\boldsymbol{p}}(-t,\,
\boldsymbol{x})$. Equation with $\mathrm{St}\, f=0$ preserves
$T$-symmetry and cannot describe kinetic
relaxation~\cite{LL10}\renewcommand{\citenumfont}[1]{#1}\cite{Zakharov-Lvov}\renewcommand{\citenumfont}[1]{S#1}, unless some
source of irreversibility is provided. Second, as we explain below,
all dangerous $O(G)$ terms vanish in the spatially homogeneous box
and in the centers spherically-symmetric halo/miniclusters, i.e.\ in
all places where we numerically observe formation of Bose stars. 
Third, our numerical results suggest that the condensation time scale
$\tau_{gr}$ is set by  $\mathrm{St}\, f$.  

A homogeneous ensemble of random waves in the box is described by
$\boldsymbol{x}$-independent real distribution
$f_{\boldsymbol{p}}$. In this case $\bar{U}=0$ and, also,  the
imaginary part of the integral in Eq.~(\ref{largeD}) vanishes, as one
can see by changing $(\boldsymbol{\xi},\, \boldsymbol{\xi}') \to
(-\boldsymbol{\xi},\, -\boldsymbol{\xi}')$. Equation (\ref{largeD}) 
takes the form 
\begin{equation}
  \label{eq:S5}
  \partial_t f  = \mbox{St}\, f \;.
\end{equation}
Thus, the time scale $\tau_{gr}$ of Bose star formation in the box is
determined  by the scattering integral explicitly.

Now, consider halo/minicluster of size $R$. We simplify
Eq.~(\ref{largeD}) by performing consistent expansion in $(mvR)^{-2} 
\ll 1$. To this end we note that $f_{\boldsymbol{p}+\Delta        \boldsymbol{p}}
(\boldsymbol{x}+\Delta\boldsymbol{x})$ can be  Taylor-expanded in
$\Delta \boldsymbol{x}\ll R$  or $\Delta \boldsymbol{p} \ll mv$ 
because the halo is almost homogeneous at   scales
$\Delta\boldsymbol{x} \ll R$, while the Wigner distribution in
Eq.~(\ref{eq:S2}) is insensitive to  $\Delta \boldsymbol{p}\ll mv$ due
to exponential falloff of the correlator at distances  exceeding the
correlation length $(mv)^{-1}$.  Performing
these Taylor expansions in the respective kinematic domains
of the integral in Eq.~(\ref{largeD}), one finds,
\begin{equation}   \label{eq:Svlasov}
  \partial_0 f + \boldsymbol{\nabla}_p h \boldsymbol{\nabla}_x f  -
    \boldsymbol{\nabla}_x  h\boldsymbol{\nabla}_{p} f  + S_3\approx
    \mathrm{St}\, f\,,
\end{equation}
where $S_3 = (m/24)\, \partial_{i} \partial_{j}
\partial_{k} \bar{U}\, \partial_{p_i}\partial_{p_j} \partial_{p_k} f$
reflects uncertainty of order $|\boldsymbol{p}|^{-1}$ in particle's
position, while 
\begin{equation} 
  \label{eq:Sh}
  h_{\boldsymbol{p}}(\boldsymbol{x})  = \frac{\boldsymbol{p}^2}{2m} +
  m\bar{U}(\boldsymbol{x}) - \frac{4\pi G m^2}{(2\pi)^3} \int \frac{d^3
  \boldsymbol{k}}{\boldsymbol{k}^2} \,
  f_{\boldsymbol{p}-\boldsymbol{k}}(\boldsymbol{x})
\end{equation}
is one-particle Hamiltonian with gravitational mass defect in the
third term. To the leading order Eqs.~\eqref{eq:Svlasov},
(\ref{eq:Sh})  reproduce the celebrated Vlasov equation. They also
include  two corrections suppressed as $(mvR)^{-2}$:  $S_3$ and 
the last term in Eq.~(\ref{eq:Sh}). We ignored
corrections  with relative suppression $(mvR)^{-4}$, as they are
smaller  than $\mathrm{St}\, f \sim f/\tau_{gr} \sim \partial_t f
/(mvR)^3$, see discussion in Sec.~6. Note that kinetic expansion of
$T$-odd $O(G)$ terms in Eq.~(\ref{largeD}) goes in powers of
$(mvR)^{-2}$, as these terms change sign under $\boldsymbol{p}\to
-\boldsymbol{p}$ reflections. 

Equation~(\ref{eq:Svlasov}) describes evolution of halos/miniclusters
in the regime~\eqref{eq:tau}. Its largest second 
and third terms vanish if the distribution is virialized i.e.\ given
by an arbitrary function of one-particle conserved quantities,
e.g.\ $f = f(h_{\boldsymbol{p}}(\boldsymbol{x}))$. Thus, these terms
describe fast collisionless mixing rather
than thermalization. Small 
kinetic correction $S_3$ takes effect at parametrically larger
time scales. It conserves spatial densities of mass, energy and
momentum at every $\boldsymbol{x}$, $\int d^{3}\boldsymbol{p} \, 
(\alpha + \beta_i p^i +  \gamma p^2/2m)\, S_3=0$, and therefore does
not induce transport in the $\boldsymbol{x}$-space. Besides,
this term becomes negligibly small in the centers of
spherically-symmetric miniclusters  $r\lesssim (mv)^{-1}$ because
$\partial^{3}\bar{U}(r)=0$ at $r= 0$. Thus, it cannot drive kinetic
relaxation in the halo centers where we numerically observe formation
of Bose stars. This leaves~$\mathrm{St}\, f$ as the only source of condensation. 

So, it is natural to expect that the time $\tau_{gr}$ of
Bose star formation inside the virialized halo/minicluster is
determined  by the scattering integral. Our numerical results
strongly support this viewpoint.  

\subsection{S1.2. Landau scattering integral}
\label{sec:s1.2.-land-scatt}
Recall that the collision integral $\mathrm{St}\, f$ is a contribution
of the connected correlator $\langle \psi \psi^{*}U\rangle_{conn}$
into the right-hand side of Eq.~(\ref{eq:S1}). Landau
noticed~\cite{LL10}\renewcommand{\citenumfont}[1]{#1}\cite{Zakharov-Lvov}\renewcommand{\citenumfont}[1]{S#1}
that in the particular case of long-range  
interactions main fluctuations of the potential $U_{\boldsymbol{x}}$
are caused by density perturbations at large distances in Eq.~(\ref{eq:S4}), 
\begin{equation} 
  \label{eq:Lregion}
  (mv)^{-1} \ll |\boldsymbol{x'-x}| \ll R\;.
\end{equation}
Below we will confirm
this nontrivial observation and estimate related corrections. In
the region~\eqref{eq:Lregion} Eq.~(\ref{eq:S4}) admits multipole
expansion,
\begin{equation}
  \label{eq:Landauexp}
  U_{\boldsymbol{x}+\boldsymbol{\xi}/2} \approx U_{\boldsymbol{x}} +
  \frac{\xi^i}{2} \partial_i U_{\boldsymbol{x}} \qquad
  \mbox{if} \;\;\;\; |\boldsymbol{\xi}| \lesssim (mv)^{-1}\;.
\end{equation}
Importantly, Eq.~(\ref{eq:Landauexp}) is approximately valid for the
fluctuations of $U$, not just for the smooth mean potential
$\bar{U}$. Then the scattering integral in Eq.~(\ref{eq:S1}) takes the
form
\begin{equation}
  \label{eq:StfLandau}
  \mathrm{St}\, f = -\boldsymbol{\nabla}_p \boldsymbol{s}\;,
\end{equation}
where the Landau flux
\begin{equation}
  \label{eq:sLandau}
  s_i = - \frac{4\pi G m^2}{(2\pi)^3} \int d^3 \boldsymbol{x}'
  d^3\boldsymbol{p}'\; {\cal F}_{\boldsymbol{x},\,
    \boldsymbol{p}}^{\boldsymbol{x}',\, \boldsymbol{p}'} \; \partial_i
  \Delta^{-1}_{\boldsymbol{x} - \boldsymbol{x'}} 
\end{equation}
describes diffusion in the phase space. It involves the connected
correlator of four fields,
\begin{multline}
  \label{eq:Fconn}
  {\cal F} = \int d^3 \boldsymbol{\xi} d^3\boldsymbol{\xi}'  \;
  \mathrm{e}^{-i \boldsymbol{p} \boldsymbol{\xi} -
    i\boldsymbol{p}' \boldsymbol{\xi}'}\\ \times \langle \psi_{\boldsymbol{x}
  + \boldsymbol{\xi}/2} \psi^*_{\boldsymbol{x}
  - \boldsymbol{\xi}/2} \psi_{\boldsymbol{x}'
  + \boldsymbol{\xi}'/2} \psi^*_{\boldsymbol{x}'
  - \boldsymbol{\xi}'/2} \rangle_{conn}\;,
\end{multline}
where $\langle \cdot \rangle_{conn}$ is a combination of four- and
two-point functions introduced in Eq.~(\ref{eq:Fpoint}). Note that the
integrals in Eqs.~(\ref{eq:Fconn}), (\ref{eq:sLandau}) are saturated
at small $\boldsymbol{\xi} \sim p^{-1}$ and~${\boldsymbol{\xi}' \to 0}$. 

We treat ${\cal F}$ in the same way as $f$ in the previous
Section. Acting by $\partial_t$ and using Eqs.~\eqref{eq:GP2}, we
obtain dynamical equation
\begin{equation}
  \label{eq:Feq}
  \partial_0 {\cal F} + (\boldsymbol{p}\boldsymbol{\nabla}_x +
  \boldsymbol{p}' \boldsymbol{\nabla}_{x'}) {\cal F}/m = {\cal A}\;,
\end{equation}
with the right-hand side ${\cal A}_{\boldsymbol{x},\,
  \boldsymbol{p}}^{\boldsymbol{x}',\, \boldsymbol{p}'}\big|_{t}$
containing correlators of six fields $\langle \psi^2 \psi^{*2}
U\rangle$ at time $t$, 
cf.\ Eq.~(\ref{eq:S1}). Next, we compute ${\cal A}$ to the leading order in
$G$ and $(mvR)^{-1}$. Namely, we express it via the two-point
functions $f$ like in Eq.~(\ref{eq:Fpoint}), but ignore all
connected parts. We also use the dipole approximation
(\ref{eq:Landauexp}) and expand $f_{\boldsymbol{p}}(\boldsymbol{x})$ 
in small variations of $\boldsymbol{p}$ and $\boldsymbol{x}$ whenever 
possible. We find, 
\begin{equation}
  \label{eq:SA}
  {\cal A} = 4 \pi G m^2 \partial_j
  \Delta^{-1}_{\boldsymbol{x} - \boldsymbol{x'}} \left[
  f'^2\partial_{p_j} f   -
  f^2 \partial_{p'_j}
  f'   \right]\;,
\end{equation}
where $f'= f_{\boldsymbol{p}'}(\boldsymbol{x}')$ and $f =
f_{\boldsymbol{p}}(\boldsymbol{x})$. Note that ${\cal A}$ in
Eq.~(\ref{eq:SA}) vanishes at large separations $|\boldsymbol{x} -
\boldsymbol{x}'| \gtrsim R$ because either $\boldsymbol{x}$ or
$\boldsymbol{x}'$  in this case is outside of the halo.

The solution of Eq.~(\ref{eq:Feq}) is, 
\begin{equation}
  \label{eq:Fsolution}
        {\cal F} = \int_{-\infty}^t dt' \; 
        {\cal A}_{\boldsymbol{x} + \boldsymbol{p}(t'-t)/m,\,
          \boldsymbol{p}}^{\boldsymbol{x'} 
           + \boldsymbol{p}' (t'-t)/m,\; \boldsymbol{p}'} \Big|_{t'}\;,
\end{equation}
where we recalled that ${\cal F}$ vanishes in the beginning of the
process. In the region~(\ref{eq:Lregion}) this solution simplifies. 
First, the left-hand side of Eq.~(\ref{eq:Feq}) suggests  that ${\cal
  F}$ responds to the external source ${\cal A}$ at time scales
$t-t' \sim |\boldsymbol{x} - \boldsymbol{x}'|/v$. Thus, we can
limit integration in Eq.~(\ref{eq:Fsolution}) to the region
$(mv^2)^{-1} \ll t-t' \ll R/v$ corresponding to Eq.~(\ref{eq:Lregion}). Then 
all $f$'s  in the integrand can be taken at time $t$. Second,
$f' \approx f_{\boldsymbol{p}'}(\boldsymbol{x})$, $f \approx
f_{\boldsymbol{p}}(\boldsymbol{x})$ in Eq.~(\ref{eq:Fsolution})
because the shifts of their spatial arguments are small in the region
(\ref{eq:Lregion}). After these approximations $\boldsymbol{x}'$
and $t'$ enter only the argument of the Green's function in
Eq.~(\ref{eq:SA}).

Substituting Eqs.~(\ref{eq:SA}), \eqref{eq:Fsolution} into
Eq.~(\ref{eq:sLandau}), one arrives at the 
Landau flux~\cite{LL10},
\begin{equation}
  \label{eq:sfinal}
  s_i = \frac{G^2 m^4}{4\pi^2} \int d^3\boldsymbol{p}'\; 
  \Pi_{ij}(\boldsymbol{u}) \; \left[f^2 \partial_{p_j'} f' - f'^2
    \partial_{p_j} f \right]\;,
\end{equation}
where $\boldsymbol{u} = (\boldsymbol{p} - \boldsymbol{p}')/m$ is the
relative velocity of the interacting waves and we denote $f\equiv
f_{\boldsymbol{p}}(\boldsymbol{x})$,  $f' \equiv
f_{\boldsymbol{p}'}(\boldsymbol{x})$,
\begin{equation}
  \label{eq:Piini}
  \Pi_{ij} = 8\pi \int dt'' \,  d^3 \boldsymbol{y}\;   \partial_i
  \Delta^{-1}_{\boldsymbol{y}} \partial_j\Delta^{-1}_{\boldsymbol{y} +
    \boldsymbol{u} t''}\;,
\end{equation}
and $\boldsymbol{y} \equiv \boldsymbol{x}-\boldsymbol{x}'$, $t''
\equiv t' - t$.

The last step is to compute $\Pi_{ij}$ in
Eq.~(\ref{eq:Piini}). Spherical symmetry gives $\Pi_{ij} =
\Pi_1(u) \delta_{ij} + \Pi_2(u) u^i u^j$. Besides,
\begin{align} 
  \label{eq:Piprop}
  & u^j \Pi_{ij} = 8 \pi\int d^3 \boldsymbol{y}\; 
  \Delta^{-1}_{\boldsymbol{y}} \, \partial_i\Delta^{-1}_{\boldsymbol{y}} =
  0\;,\\
  \notag
  & \delta^{ij} \Pi_{ij} = -\frac{2}{|\boldsymbol{u}|}
  \int\limits_{-R/v}^{-(mv^2)^{-1}} \frac{dt''}{t''} =
  \frac{2}{|\boldsymbol{u}|}  \, \log(mvR)\;,
\end{align}
where in the first line we used $u^j\partial_j = \partial_{t''}$.  In
the second line we integrated by parts, then substituted
${\Delta^{-1}_{\boldsymbol{x}} =  -(4\pi  |\boldsymbol{x}|)^{-1}}$. This
leaves divergent integral over $t''$ which is logarithmically sensitive
to the boundaries of the Landau region $(mv^{2})^{-1} \ll |t''| \ll
R/v$. We obtain,
\begin{equation}
  \label{eq:Pifinal}
  \Pi_{ij} = \Lambda\; \left(u^2 \delta_{ij} - u_i u_j\right)/u^3\;,
\end{equation}
where $\Lambda \equiv \log(mvR)$ and ${u \equiv |\boldsymbol{u}|}$. 

Expressions (\ref{eq:StfLandau}), (\ref{eq:sfinal}),
(\ref{eq:Pifinal}) give Landau scattering integral in its canonical
form. Note that the relaxation time estimate from the main text can be
drawn  directly from these expressions,
cf.~\renewcommand{\citenumfont}[1]{#1}\cite{Tkachev:1991ka}\renewcommand{\citenumfont}[1]{S#1}. Indeed, trading the $p$-derivatives for
$(mv)^{-1}$, integral over momentum for ${V_p \sim 4 \pi m^3  v^3/3}$,
using $u \sim v\sqrt{2}$ and $f\sim (2\pi)^3 n/V_p$, one obtains
${\mathrm{St}\, f \sim f/\tau_{gr}}$, where $\tau_{gr}$ is given by
Eq.~\eqref{eq:tgra3}. 

Now, let us discuss corrections to the Landau integral related to wave
scattering at very small and very large distances,  $|\boldsymbol{x} -
\boldsymbol{x'}| \lesssim (mv)^{-1}$ and $|\boldsymbol{x}-\boldsymbol{x}'|
\sim R$. In the former case one can Fourier-transform
Eq.~(\ref{eq:S4}) introducing momentum transfer $\boldsymbol{q}$
conjugate to $\boldsymbol{x}-\boldsymbol{x}'$. Scattering processes
with  $|\boldsymbol{q}|  \gg mv$ do not occur because waves of such high
momenta are absent in the ensemble. Thus, true scattering integral has a
cutoff at ${q^{-1} \sim |\boldsymbol{x} -   \boldsymbol{x}'| \gtrsim
  (mv)^{-1}}$ that regularizes logarithmic divergence at small $|t''|$ in
Eq.~(\ref{eq:Piprop}). A similar cutoff at  $v|t''| \sim
|\boldsymbol{x}-\boldsymbol{x}'| \gtrsim R$ is related to the fact
that matter is absent outside of the halo/minicluster. Clearly, uncontrolled
contributions from the regions near the cutoffs are comparable to the
Landau scattering integral, but have $O(1)$ constants in front instead
of the Coulomb  logarithm ${\Lambda \equiv   \log(mvR)}$. Thus, Landau
approximation keeps terms proportional to $\Lambda \gg 1$ 
and ignores relative $O(1)$ corrections. In Sec.~5 we treat these
corrections heuristically, by changing $\Lambda \to \Lambda + a$,
where $a \sim O(1)$. In Fig.~\ref{fig:sections1}f we demonstrate that
Landau equation with $a\approx 5$ correctly  describes evolution of
the distribution  function. 

It is worth noting that the traditional Boltzmann integral also has
logarithmic accuracy in $\Lambda \gg 1$ in the case of isolated
halo/minicluster. Indeed, it diverges 
logarithmically~\renewcommand{\citenumfont}[1]{#1}\cite{Zakharov-Lvov}\renewcommand{\citenumfont}[1]{S#1} at small momentum  transfers 
$\boldsymbol{q}\to 0$ due to long-range nature of gravitational
interactions. Unsuppressed contributions at  $|\boldsymbol{q}| \sim
R^{-1}$ correspond to interactions at distances of order of the 
structure size. Imposing cutoff at these scales, one recovers finite
result up to uncontrolled $O(1)$ terms. This situation is
drastically different from the case of electromagnetic plasma where
the Coulomb interactions disappear at distances larger than the
microscopic  Debye length $\lambda_D \ll R$ which regularizes the
Boltzmann integral~\cite{LL10}.

\subsection{S1.3. Energy distribution}
Let us argue that Eq.~\eqref{eq:1} gives distribution of bosons over
energies in the kinetic regime~\eqref{eq:tau}: $F(t,\, \omega) \approx
dN/d\omega$. On the one hand, this expression involves average over
large time interval $\tau_1  \gg (mv^2)^{-1}$ which is equivalent to
the ensemble average due to ergodic hypothesis: $F \approx  \langle
F\rangle$. On the other hand, we consider regime where at  time scales
$\tau_1 \ll \tau_{gr}$ the mean potential $\bar{U}(\boldsymbol{x})$ is
almost static and the field $\psi$ evolves in this potential. Thus,
${\psi \approx \sum_nc_n\, \psi_n(\boldsymbol{x})\,
  \mathrm{e}^{-i\omega_n   t}}$, where $\psi_n(\boldsymbol{x})$ are
the  eigenstates of $\bar{U}$ with almost time-independent
eigenenergies $\omega_n$, while $|c_n|^2$ are the occupation
numbers. Substituting this form into Eq.~\eqref{eq:1}, we obtain,
\begin{equation}
  \label{eq:SFresult}
  F \approx \sum_n \langle |c_n|^2 \rangle\;  \delta_{\tau_1}(\omega - \omega_n)\;,
  \;\; 
\end{equation}
where $\delta_{\tau_1} = (\tau_1/2\sqrt{\pi})\, \mathrm{e}^{-
  \tau_1^2(\omega- \omega_1)^2/4}$ is a sharply-peaked
function indistinguishable from $\delta(\omega - \omega_n)$ at
energy resolution  ${\Delta\omega \gtrsim \tau_1^{-1}}$.  Thus, for chosen
values ${\tau_1^{-1} \ll  mv^2 \sim \omega}$ the function $F$ gives
distribution of particles over energies $dN/d\omega$.

\subsection{S1.4. Homogeneous isotropic ensemble}
Finally, we derive kinetic equation~\eqref{eq:2} for the 
homogeneous spherically-symmetric ensemble of random classical waves
evolving in the box of size $L$, cf.~\renewcommand{\citenumfont}[1]{#1}\cite{Zakharov-Lvov}\renewcommand{\citenumfont}[1]{S#1}. In this
case $f_p$ depends only 
on ${p = |\boldsymbol{p}|}$ and the Landau flux $\boldsymbol{s}$ is
collinear with momentum,  $\boldsymbol{ps} = ps$. Equations
(\ref{eq:S5}), (\ref{eq:StfLandau}), (\ref{eq:sfinal}) give,
\begin{equation}
  \label{eq:spheq}
  \partial_t f = \frac{2 G^2 m^5}{\pi p^2}\,  (\Lambda + a) \, \partial_p
  \left[ A\partial_p f+ Bf^2\right]\;,
\end{equation}
where we computed the angular integrals and introduced ${A =
  \int_0^{\infty} p'dp' \, \min(p^3,\, p'^3) \,f'^2/3p}$, ${B = 
\int_0^p p'^2 dp' \, f'}$. In Eq.~(\ref{eq:spheq}) we included
the heuristic parameter $a$ from Sec.~S1.2. 

Now, we express $f_p$ in terms of the energy distribution
$F(\omega)$. Since $dN = F d\omega = 4\pi p^2 L^3 f_p
dp/(2\pi)^3$, we have,
\begin{equation}
  \notag
  F(\omega) = m L^3  p f_p/2\pi^2 \;, \qquad \omega = p^2/2m\;.
\end{equation}
Substituting this expression into Eq.~(\ref{eq:spheq}), we find, 
\begin{multline}
  \label{eq:Fnew}
  \partial_t F = \frac{2}{\pi} G^2 m^3 (\Lambda + a)  \\ \times
  \partial_\omega  \left[ A \partial_\omega F + ( 2\pi^2 BF/mL^3 -
    A) F/2\omega \right] \;,
\end{multline}
where 
\begin{align}
  \notag
  &A(\omega) = \frac{4\pi^4}{3mL^6} \int_0^\infty
  d\omega'\, \mathrm{min}^{3/2}(\omega, \omega')
  \, F'^2/\omega' \omega^{1/2} \;,\\
  \notag
  &B(\omega) = \frac{2\pi^2}{L^3} \int_0^\omega d\omega'\,  F'
\end{align}
are the same quantities as in Eq.~(\ref{eq:spheq}).

The final step is to introduce dimensionless energy $\tilde{\omega} =
\omega/mv_0^2$ and rescaled distribution $\tilde{F} = mv_0^2 F/N$
normalized to unity: $\int_0^{\infty} d\tilde{\omega}\, \tilde{F} =
1$. We obtain Eq.~\eqref{eq:2} with ${\tilde{A} = m^2 v_0^2
A/ 4 \pi^4 n^2}$, $\tilde{B} = B/2\pi^2 n$. Note that the kinetic time
scale $\tau_0\sim \tau_{gr}$ is explicit in Eq.~\eqref{eq:2} because
all physical parameters are scaled out of this equation.

\section{S2. Numerical method}
\label{sec:numerical-method}
We perform numerical studies using 6-th order pseudo-spectral
operator-splitting method~\cite{Yoshida:1990zz} which is unitary,
stable, $T$-symmetric, symplectic, and therefore exceptionally 
suitable for long statistical simulations. In what 
follows we explain the method, illustrate its properties and
estimate the numerical errors. 

Recall that general solution to the Schr\"odinger 
equation~\eqref{eq:GP2} has the form $\psi(t+\Delta t)
= \hat{\cal U} \psi(t)$, where 
\begin{equation}
  \notag
  \hat{\cal U} = T\exp \left\{-i \int_t^{t+\Delta t}
  dt'\, \left[\frac{\hat{\boldsymbol{p}}^2}{2m}+ mU(t',\,
    \boldsymbol{x})\right] \right\}  \;,
\end{equation}
is a quantum propagator and $\hat{\boldsymbol{p}} \equiv
-i\boldsymbol{\nabla}_x$. The 
method of~\cite{Yoshida:1990zz} replaces this propagator with  
discrete formula
\begin{multline}
  \label{symp}
  \hat{\cal U} = \prod_{\alpha =1}^8  \mathrm{e}^{-i m d_\alpha\Delta
    t U_\alpha(\boldsymbol{x}) } \mathrm{e}^{-ic_{\alpha}
    \Delta t \hat{\boldsymbol{p}}^2/2m}  
   + O(\Delta t^7)\;,
\end{multline}
where the product is ordered right-to-left, its parameters $c_\alpha$,
$d_\alpha$ are given in Table~\ref{tab:parameters}, and the potentials
$U_\beta$ are computed 
using the ``current'' field, i.e.\ $\psi(t)$ multiplied by all
operators with $\alpha \leq \beta$. Geometrically,
Eq.~(\ref{symp}) breaks the time interval $\Delta t$ into two sets of
sub-intervals $\{c_\alpha \Delta t,\, d_\alpha \Delta t\}$, where each
set is symmetric with respect to the central point $t +  \Delta
t/2$. ``Kinetic'' and ``potential'' propagators are used  for the $c$-
and $d$-sub-intervals, respectively. We stress that Eq.~(\ref{symp}) is valid
for our time-dependent potential $U(t,\, \boldsymbol{x})$.
\begin{table}[t]
\centerline{\begin{tabular}{r|c|c|c|c}
    \hline
    $\alpha$      & $1$     & $2$             & $3$            & $4$\\\hline
    $c_\alpha$ & $w_3/2$ & $(w_2+w_3)/2$ & $(w_1+w_2)/2$ & $(w_0+w_1)/2$\\
    $d_\alpha$   & $w_3$ & $w_2$         & $w_1$         & $w_0$\\\hline
\end{tabular}}

\begin{align*}
  &c_{9-\alpha} = c_\alpha \;, \;\;\; d_{8-\alpha} = d_\alpha\;, \;\;\; d_8 =
  0\;, \;\;\; \sum c_\alpha = \sum d_\alpha = 1\;,&\\
  &\begin{aligned}
    & w_0 = 1-2(w_1+w_2+w_3)\;,  & w_1 = -1.17767998417887\;,\\
    & w_2 = 0.235573213359359\;, &w_3 = 0.784513610477560\;.
  \end{aligned}&
\end{align*}
\caption{Parameters in Eq.~(\ref{symp}).\label{tab:parameters}}
\end{table}

Numerical application of Eq.~(\ref{symp}) is straightforward. One
introduces cubic uniform lattice with $N_x^3$ sites at ${\boldsymbol{x}   =
    \boldsymbol{n}_x \Delta x}$,  where ${\Delta x\equiv L/N_x}$ and $0 \leq
n_x^i < N_x$.  Typically, we  use $N_x=128$ or $256$, and
switch to $N_x =   512$ for resolution tests. We store the values
of  the fields $\psi(\boldsymbol{x})$, $U(\boldsymbol{x})$ at the
lattice sites.

Time evolution of $\psi(\boldsymbol{x})$ is calculated by
sequentially acting with operators in Eq.~(\ref{symp}). We  start by
performing the Fourier transform,   
\begin{equation}
  \label{psip}
  \psi_{\boldsymbol{p}} =
  \Delta x^3\sum_{\boldsymbol{x}} \psi(\boldsymbol{x})\;
  \mathrm{e}^{-i\boldsymbol{p} \boldsymbol{x}}\;,
\end{equation}
where the momenta $\boldsymbol{p} = 2\pi \boldsymbol{n}_p/L$ are
discrete, $-N_x/2 < n_p^i \leq N_x/2$, due to periodic boundary
conditions ${\psi(\boldsymbol{x}+L\boldsymbol{n'}) =
  \psi(\boldsymbol{x})}$. We apply Eq.~(\ref{psip}) using FFT
algorithm with GPU acceleration~\cite{cuda}. After that multiplication
by the rightmost operator in Eq.~(\ref{symp}) corresponds to phase
rotation ${\psi_{\boldsymbol{p}} \to   \psi_{\boldsymbol{p}} \,
  \mathrm{e}^{-i    c_1 \Delta t\,     \boldsymbol{p}^2
    /2m}}$. We return $\psi$ to the coordinate representation with
the inverse Fourier transform.  Next, we solve the Poisson
equation~\eqref{eq:GP2} in the momentum space,    
\begin{equation}
  \label{Usolution}
  U_{1,\, \boldsymbol{q}} = -4\pi m G
  |\psi^2|_{\boldsymbol{q}}/\boldsymbol{q}^2\;, \qquad
  U_{1,\, \boldsymbol{q}=\boldsymbol{0}} = 0\;,
\end{equation}
where $|\psi^2|_{\boldsymbol{q}}$ is the Fourier image of
$(|\psi(\boldsymbol{x})|^2-n)$ calculated via
Eq.~(\ref{psip}). Finding $U_1(\boldsymbol{x})$ from the  inverse
Fourier transform, we act by the ``potential'' propagator,
i.e.\  change ${\psi(\boldsymbol{x}) \to   \psi(\boldsymbol{x})\,
  \mathrm{e}^{-imd_1   \Delta t \,U_1(\boldsymbol{x})}}$. We  continue to act by
operators in Eq.~(\ref{symp}) on $\psi_{\boldsymbol{p}}$ or
$\psi(\boldsymbol{x})$ until $\psi(t+\Delta t)$ is found. 

Let us illustrate advantages of the above numerical method and
estimate related errors.

\begin{figure}[t!]
  \centerline{\includegraphics{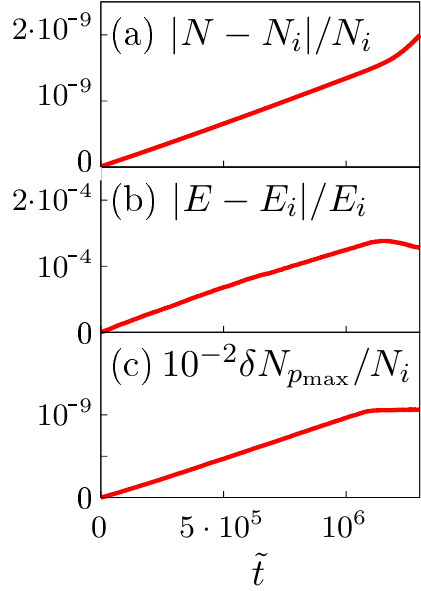}\includegraphics{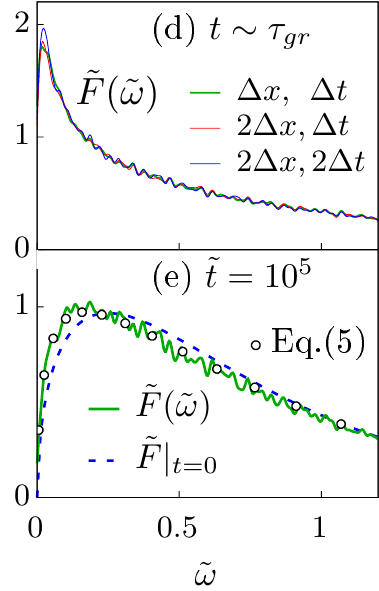}}

  \vspace{2mm}
  \centerline{\includegraphics{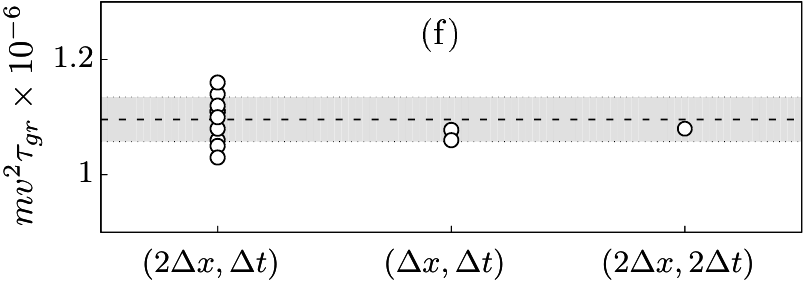}}
  \caption{Tests of the numerical solution in Fig.~\ref{fig:sections1}
    which is computed on $256^3$ lattice with $\Delta\tilde{t} \sim
    1.5,$ $\Delta \tilde{x}\approx 0.5$. (a), (b)~Drift of the particle
    number $N$ and energy $E$ from their initial values $N_i$,
    $E_i$. (c)~Discretization error estimate.  (d)~Spectra 
    computed at $\tilde{t} = {1.05\cdot 10^6}\sim \tilde{\tau}_{gr}$ 
    with different space and time resolutions. (e)~Spectrum at $\tilde{t} 
    = 10^5$, $\Delta \tilde{t} \sim 0.5$ versus
    Eq.~\eqref{eq:2}. (f)~The time of Bose star formation  
    $\tau_{gr}$ at different space and time resolutions, where $\Delta
    t$ and $\Delta x$ are the parameters of the 
    original solution.  \label{fig:long_run}}
\end{figure}

{\bf 1. Exact conservation of the particle number.} One notes that
the Fourier transform (\ref{psip}) and subsequent multiplication by
phases do not change the particle number 
\begin{equation}
  \label{NNnew}
  N = \Delta x^3 \sum_{\boldsymbol{x}}
  |\psi(\boldsymbol{x})|^2 =
  L^{-3} \sum_{\boldsymbol{p}} |\psi_{\boldsymbol{p}}|^2\;,
\end{equation}
which therefore conserves up to round-off errors. Hence, one expects
relative change $\Delta   N/N   \lesssim 2^{-52} \,    N_x^{3/2} \sim
10^{-12}$ during one time step, where we stick to
double-precision numbers and $N_{x}  =256$. In practice we observe
even better conservation: the total drift of $N(t)/N_i$ does not exceed
$10^{-8}$ even for our longest runs with $10^{6}$ time steps, see
Fig.~\ref{fig:long_run}a. Thus,  particles do not appear from nowhere
in our numerical system.  

{\bf 2. Time resolution.} We use adaptive step size $\Delta t$ which
is chosen to enforce conservation of energy 
\begin{equation}
  \notag
  E = L^{-3} \sum_{\boldsymbol{p}} \frac{\boldsymbol{p}^2}{2m} \,
  |\psi_{\boldsymbol{p}}|^2 +  \Delta
    x^3\sum_{\boldsymbol{x}} \frac{m}{2}\,U(\boldsymbol{x})\,
    |\psi(\boldsymbol{x})|^2
\end{equation}
between $t$ and $t+\Delta t$ with accuracy ${\Delta E/E \leq 2
  \cdot 10^{-10}}$. This corresponds to values ${\Delta \tilde{t} \sim 0.1
  \div 1.5}$ in our $O(\Delta  t^6)$ scheme. As a consequence,
relative non-conservation of energy in our solutions never exceeds
$10^{-3}$, see Fig.~\ref{fig:long_run}b.  

Time resolution of the solutions is proportional to $\Delta E/E$. We
estimate it explicitly by performing  two steps $\Delta t/2$ instead  of one 
step $\Delta t$ for a given $\psi(t,\, \boldsymbol{x})$. Changes in
${\psi(t+\Delta  t,\,   \boldsymbol{x})}/n^{1/2}$ are of
order~${10^{-6} \sim 10^{3} \, \Delta E/E}$. Below we show, by
drastically increasing and decreasing the time step, that this 
resolution is good enough to make numerical results independent
of~$\Delta t$. 

Note that  evolution of the waves with high momentum
$\boldsymbol{p}$ is correctly described by Eq.~(\ref{symp}) that
exactly reproduces fast oscillations with frequency $\boldsymbol{p}^2
/2m$. Errors in $\Delta t$ stem from the ``potential'' propagators
with bounded~${\tilde{U}   \Delta   \tilde{t} \ll 10^{-2}}$.

{\bf 3. Discretization errors.} To estimate
effects due to spatial discretization, we consider inverse Fourier
transform in the box, 
\begin{equation}
  \label{Finverse}
  \psi(\boldsymbol{x}) = L^{-3} \sum_{\boldsymbol{p}} \psi_{\boldsymbol{p}}\;
  \mathrm{e}^{i\boldsymbol{p} \boldsymbol{x}}\;.
\end{equation}
This representation would be exact if the sum over ${\boldsymbol{p} =
2 \pi \boldsymbol{n}_p/L}$  was infinite. But in fact, it has a cutoff
at ${p_{max} = \pi/\Delta   x}$. Hence, discretization errors are
determined by  
$|\psi_{p_{max}}|$. One estimates error in the particle
number, Eq.~(\ref{NNnew}), as ${\delta N_{p_{max}} \sim
  p_{max}^3\,|\psi_{p_{max}}|^2 
  /6\pi^2}$. Relative changes in Eqs.~(\ref{Usolution}),
(\ref{symp}) due to $p_{max}$ are proportional to this quantity.
Thus, we can relate the discretization errors to~$\delta 
N_{p_{max}}/N$. 

In Fig.~\ref{fig:long_run}c we demonstrate that solutions with ${N_x
  =256}$  have $\delta  N_{p_{max}}/N \lesssim 10^{-7}$.  We
independently estimated the discretization errors by performing one
time step on $256^3$ and $512^3$ lattices.  Changes in $\psi(t+\Delta
t,\, {x})/n^{1/2}$ were smaller than $10^{-9}$, which makes 
spatial resolution at one time step of order $10^{-2} \, \delta N_{p_{max}}/N$. When
performing the calculations, we never allowed this combination to grow
above $10^{-8}$.

One explains exceptionally good spatial resolution using the spectrum
at $t \sim \tau_{gr}$ in Fig.~\ref{fig:sections1}f. The central part
of this spectrum is close to the power-law $F  \propto
\omega^{-1/2}$, which changes to
exponential falloff in the region $\tilde{\omega} \gtrsim 
1.5$. As a consequence, $F \propto p |\psi_p|^2$ is exponentially 
small at the cutoff  frequency ${\omega_{\max} =  
  p_{max}^2/2}$, where $\tilde{\omega}_{max}\approx 21$ for the
solution in Fig.~\ref{fig:sections1}f. Note that with time
the central power-law  region of the distribution spreads  to higher
and lower~$\omega$. Eventually, it should  hit the cutoff
frequency. However, way before that it reaches 
$\omega\approx 0$ at $t=\tau_{gr}$, and the Bose star is formed. Thus,
our numerical solutions correctly describe the initial stages of
thermalization, and  related discretization errors are
exponentially small  at~$t \lesssim \tau_{gr}$.

We remark that spatial resolution at smaller $\tilde{L}$ and the same
$N_x=256$ is orders of magnitude better than in 
Fig.~\ref{fig:long_run}.  We will use this property below. 

{\bf 4. Stability.} In general, good resolutions in time and space are
not sufficient to reproduce correct statistical properties of the wave
ensemble at large time scales $\tilde{t}\sim 10^6$. The main property
for that is stability, i.e.\ absence of exponentially growing or
decaying numerical Lyapunov exponents in addition to the ones that are
already present in the continuous system. It is well-known that
pseudo-spectral methods like ours are exceptionally  stable because
they exactly conserve symplectic form~\cite{Yoshida:1990zz} and as
a consequence, the phase volume  
\begin{equation}
  \notag
  \Omega = \prod_{\boldsymbol{x}}d\psi_{\boldsymbol{x}}
  d\psi_{\boldsymbol{x}}^* = \prod_{\boldsymbol{p}}
  \frac{d\psi_{\boldsymbol{p}} d\psi_{\boldsymbol{p}}^*}{(L\Delta
    x)^3}\;, \;\;\; \Omega(t+\Delta t) = \Omega(t)\;,
\end{equation}
where $\psi$ and $\psi^*$ are the canonical variables in field
theory, while $d\psi(t,\, \boldsymbol{x})$ is the difference of
two close solutions within the statistical ensemble. Note that
conservation  of  $\Omega$ under Eq.~(\ref{symp}) can be
demonstrated directly by using $2d\psi_{\boldsymbol{x}}
d\psi^*_{\boldsymbol{x}} = d|\psi_{\boldsymbol{x}}|^2\,  d\mathrm{arg}\, 
\psi_{\boldsymbol{x}}$ and recalling that $U$ depends only on
$|\psi_{\boldsymbol{x}'}|^2$. 

The above property means that Eq.~(\ref{symp}) replaces 
continuous Hamiltonian evolution with symplectic discrete analog, where the
Liouville theorem remains satisfied. Practice shows that this
precludes appearance of numerical instabilities or exponentially
growing errors in conserved quantities, see~\cite{Yoshida:1990zz} and
references therein.  

\begin{figure}[t!]
  \begin{flushleft}
    \includegraphics{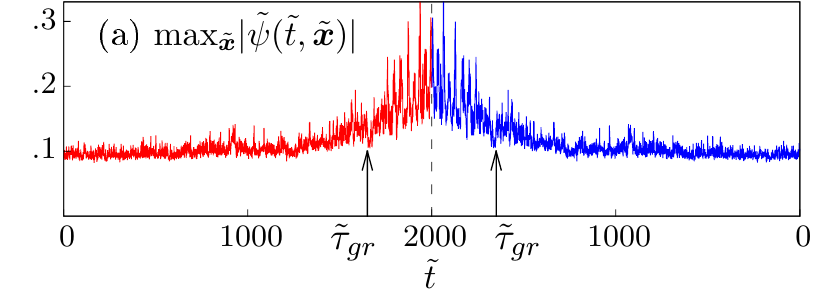}
  \end{flushleft}
  \begin{flushright}
    \includegraphics{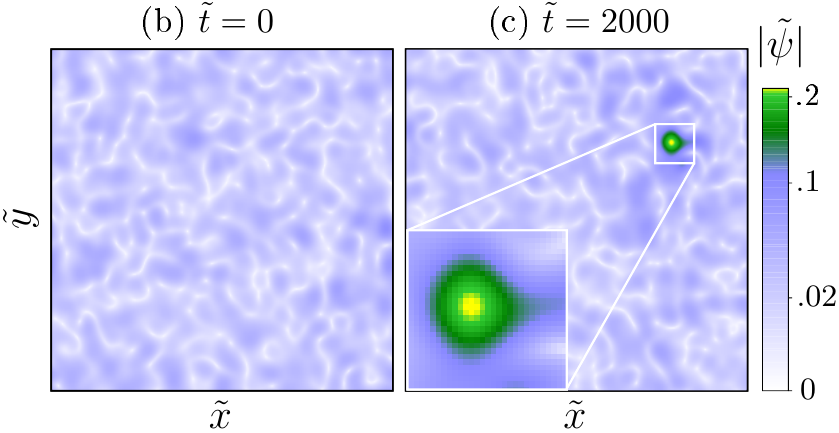}
  \end{flushright}
  \begin{flushleft}
    \includegraphics{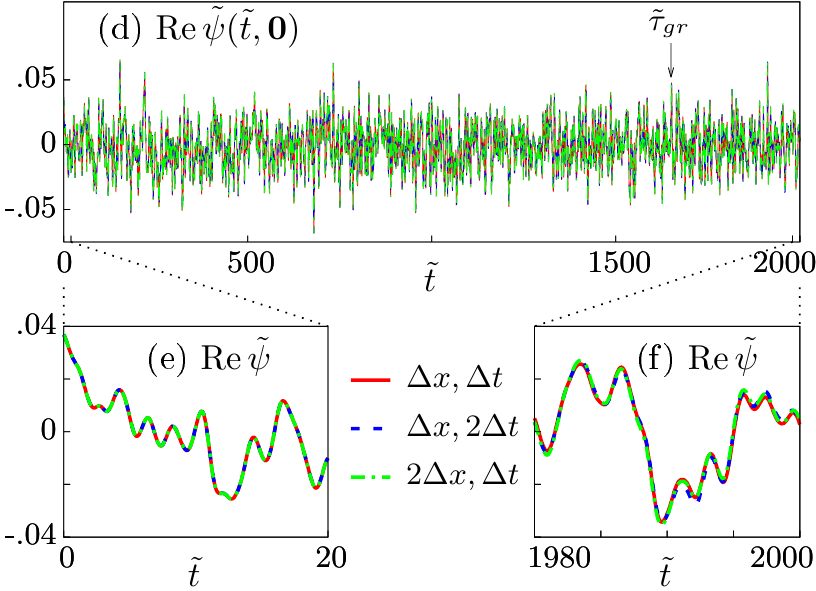}
  \end{flushleft}

  \caption{Tests of the numerical solution with $\tilde{L} = 14\pi$,
    ${\tilde{N} = 55}$ and initial distribution
    $|\psi_{\boldsymbol{p}}|^2 \propto \delta(|\boldsymbol{p}|-mv_0)$;
    in this case $\tilde{\tau}_{gr} \approx 1650$. The reference solution is
    computed on $256^3$ lattice with uniform time step $\Delta
    \tilde{t} = 0.01$ and $\Delta \tilde{x}\approx 0.2$. (a)~Maximum of
    $|\tilde{\psi}(\tilde{\boldsymbol{x}})|$ over the box as a
    function of time. Two parts of the graph correspond to direct
    evolution to $\tilde{t}=2000$ followed by evolution with steps
    $(-\Delta \tilde{t})$ to $\tilde{t}=0$. (b),~(c)~Sections $z = \mbox{const}$
    of the solution at $\tilde{t}=0$ and $\tilde{t}=2000$. (d)~Time
    evolution of $\mathrm{Re}\, \psi(\boldsymbol{x}=0)$. Coinciding lines of different color represent
    solutions with different resolutions in space and time. 
    Intervals $0 < \tilde{t}  < 20$ and $1980< \tilde{t} < 2000$ are  
    amplified in the insets~(e),~(f).\label{fig:short_run}}
\end{figure}
{\bf 5. Independence of the lattice parameters.}
Let us explicitly verify that statistical properties of our
system are not sensitive to $\Delta x$ and $\Delta t$. In
Fig.~\ref{fig:long_run}d we demonstrate that energy distributions
$F(\omega)$ at ${t \sim  \tau_{gr}}$ coincide for solutions computed
with different $\Delta t$, $\Delta x$, and the same initial
distribution at  $t=0$. Similarly, Fig.~\ref{fig:long_run}f  shows
that changes in $\tau_{gr}$ due to different choices of $\Delta t$ and
$\Delta x$ are smaller than statistical fluctuations of this quantity
related to random initial state; the standard deviation of $\tau_{gr}$
is shown with the gray strip on this graph. Thus,  our numerical
method is stable, while spatial and time resolutions are small enough
to make numerical results insensitive to lattice parameters.

{\bf 6. Exact time-reversibility.} Formation of Bose stars is an
irreversible process. These objects appear in all of our $\sim 10^3$
solutions and never disappear, their masses grow monotonically with
time. Let us argue that this effect is physical i.e.\  unrelated to 
accumulation of numerical errors. To this end we note
that Eq.~(\ref{symp}) has exact time-reversal symmetry: applying it
with time step ${-\Delta t<0}$, one reproduces all stages of the
algorithm in reverse order. Moreover, the latter operation is
equivalent to substitution $\psi\to \psi^*$ followed by evolution with positive
$\Delta t$. Thus, irreversible numerical errors can be related only
to small round-off effects.  

Figure~\ref{fig:short_run}a explicitly demonstrates time-reversibility
of our numerical code. It shows maximal value of
$|\psi(\boldsymbol{x})|$ within the box as a function of time for the
solution which forms a Bose star at $\tilde{\tau}_{gr} 
\approx 1650$, see Figs.~\ref{fig:short_run}b,c. At $t>\tau_{gr}$
the maximal density $|\psi(\boldsymbol{x})|^2$ corresponds to the Bose
star center, it increases with time due to growth of this object. At 
$\tilde{t}=2000$ we flip $\Delta t \to -\Delta t$ and continue
simulation until $t=0$ is reached, again. Although all
accumulated numerical errors remain in the system, our solution in
Fig.~\ref{fig:short_run}a correctly describes time-reversed
dynamics: density of the Bose star decreases, and this object
dissolves in the wave ensemble. We stress that the second part of
Fig.~\ref{fig:short_run}a is statistically unprobable, we never
observed anything like that in the ordinary simulation runs.  

We conclude that small round-off errors in our code are not
responsible for the irreversible processes of Bose star formation. 

{\bf 7. Continuum limit.}
One usually does not consider continuum limit of solutions in
large-scale statistical simulations such as ours or e.g.\ $N$-body
simulation. Indeed, evolution of interacting  classical systems is
intrinsically unstable. It leads to exponentially fast separation of
solutions with different $\Delta x$ and $\Delta t$ and therefore
precludes direct evaluation of the limit $\Delta t$, $\Delta x \to 0$
even at moderately large~$t$. Nevertheless, numerical errors have
negligible effect on the results of such simulations. Indeed, they
mimic small statistical fluctuations in the random system and
therefore do 
not show up in  averages. In Fig.~\ref{fig:long_run}d,f we explicitly
verified this property~---  demonstrated that $F(\omega)$ and
$\tau_{gr}$ are not sensitive to $\Delta t$ and $\Delta x$. This is
sufficient to show that the statistical simulations are correct. 

Nevertheless, our high-precision numerical code gives a unique
opportunity to demonstrate explicitly that Bose star formation occurs
in  the continuum limit. To this end we consider the solution in
Fig.~\ref{fig:short_run} which is computed with the best possible time and 
space resolutions of order $10^{-13}$. This means that the numerical
solutions with different $\Delta x$ and $\Delta t$ coincide at every
lattice point with relative precision $10^{-13}$ at the very  first
time step, see Figs.~\ref{fig:short_run}d,e. The difference between these
solutions, however, grows exponentially with time, becomes of order
$10^{-3}$ at $t = \tau_{gr}$ and reaches $5\%$ at
$\tilde{t}=2000$, see Fig.~\ref{fig:short_run}f. Thus,  within the
interval $0 < \tilde{t} < 2000$ all these solutions are close to the
continuum limit. They explicitly prove that the Bose star is formed at
$\tilde{\tau}_{gr}\approx 1650$ in the continuum system.

Note that the solution in  Fig.~\ref{fig:short_run} confirms our
kinetic formula~\eqref{eq:tgra3} for $\tau_{gr}$: it is represented by
the leftmost lower square in Fig.~\ref{fig:tgra}.   Let us verify the
time  scales $\tau_{gr}$ of other solutions~--- say, the one in
Figs.~\ref{fig:sections1},~\ref{fig:long_run}. To this end we repeat
the simulation using smaller time steps ${\Delta \tilde{t}=0.25}$ and
$0.5$, and lattices $256^3$,  $512^3$. At $\tilde{t}<10^5$ all these
solutions coincide at all lattice points with relative precision
better than $10^{-2}$. Hence,  they are close to the continuous limit
within this time interval. In Fig.~\ref{fig:long_run}e we plot their
energy distribution $F(\omega)$ at $\tilde{t}=10^5$. It coincides with
the solution to the kinetic equation~\eqref{eq:2} (points in
Fig.~\ref{fig:long_run}f) thus confirming the time scale $\tau_0
\approx \tau_{gr}$ in this  equation. Recall that our numerical
solutions  with lower resolution equally well coincide with solution
to Eq.~\eqref{eq:2}, see Fig.~\ref{fig:sections1}f. Thus, $\tau_{gr}$
is correctly computed in the main text.

In this Section we demonstrated that our numerical solutions correctly
describe evolution of the gravitating field, and our results are not
sensitive to the lattice parameters. 


\end{document}